%% file: Final.tex
\documentclass[aps,reprint,amsmath,amssymb,amsfonts,superscriptaddress]{revtex4-1}

\usepackage{color}
\usepackage{graphicx}   
\usepackage{array}
\usepackage{dcolumn}    
\usepackage{bm}         
\usepackage[normalem]{ulem}

\begin{document}
\title{Exploration of Intercell Wireless Millimeter-Wave Communication in the Landscape of Intelligent Metasurfaces}

\author{Anna C. Tasolamprou}\email{atasolam@iesl.forth.gr}
\affiliation{Institute of Electronic Structure and Laser, Foundation for Research and
Technology Hellas, N. Plastira 100, 70013 Heraklion, Crete, Greece}
\author{Alexandros Pitilakis}
\affiliation{Institute of Electronic Structure and Laser, Foundation for Research and
Technology Hellas, N. Plastira 100, 70013 Heraklion, Crete, Greece}
\affiliation{Aristotle University of Thessaloniki
Department of Electrical and Computer Engineering, 54124, Thessaloniki, Greece}
\author{Sergi Abadal}
\affiliation{NaNoNetworking Center in Catalonia (N3Cat), Universitat Politecnica de Catalunya, Barcelona, Spain}
\author{Odysseas Tsilipakos}
\affiliation{Institute of Electronic Structure and Laser, Foundation for Research and
Technology Hellas, N. Plastira 100, 70013 Heraklion, Crete, Greece}
\author{Xavier Timoneda}
\affiliation{NaNoNetworking Center in Catalonia (N3Cat), Universitat Politecnica de Catalunya, Barcelona, Spain}

\author{Hamidreza Taghvaee}
\affiliation{NaNoNetworking Center in Catalonia (N3Cat), Universitat Politecnica de Catalunya, Barcelona, Spain}

\author{Mohammad Sajjad Mirmoosa}
\affiliation{Department of Electronics and Nanoengineering, Aalto University, P.O. Box 15500, FI-00076 Aalto, Finland}

\author{Fu Liu}
\affiliation{Department of Electronics and Nanoengineering, Aalto University, P.O. Box 15500, FI-00076 Aalto, Finland}

\author{Christos Liaskos}
\affiliation{Institute of Electronic Structure and Laser, Foundation for Research and
Technology Hellas, N. Plastira 100, 70013 Heraklion, Crete, Greece}

\author{Ageliki Tsioliaridou}
\affiliation{Institute of Electronic Structure and Laser, Foundation for Research and
Technology Hellas, N. Plastira 100, 70013 Heraklion, Crete, Greece}

\author{Sotiris Ioannidis}
\affiliation{Institute of Electronic Structure and Laser, Foundation for Research and
Technology Hellas, N. Plastira 100, 70013 Heraklion, Crete, Greece}

\author{Nikolaos V. Kantartzis}
\affiliation{Institute of Electronic Structure and Laser, Foundation for Research and
Technology Hellas, N. Plastira 100, 70013 Heraklion, Crete, Greece}
\affiliation{Aristotle University of Thessaloniki
Department of Electrical and Computer Engineering, 54124, Thessaloniki, Greece}

\author{Dionysios Manessis}
\affiliation{System Integration and Interconnection Technologies, Fraunhofer IZM, Berlin, Germany}

\author{Julius Georgiou}
\affiliation{Department of Electrical and Computer Engineering, University of Cyprus, Cyprus}

\author{Albert Cabellos-Aparicio}
\affiliation{NaNoNetworking Center in Catalonia (N3Cat), Universitat Politecnica de Catalunya, Barcelona, Spain}

\author{Eduard Alarcon}
\affiliation{NaNoNetworking Center in Catalonia (N3Cat), Universitat Politecnica de Catalunya, Barcelona, Spain}

\author{Andreas Pitsillides}
\affiliation{Department of Computer Science, University of Cyprus, Cyprus}

\author{Ian F. Akyildiz}
\affiliation{Department of Computer Science, University of Cyprus, Cyprus}
\affiliation{Georgia Institute of Technology, School of Electrical and Computer Engineering, USA}

\author{Sergei A. Tretyakov}
\affiliation{Department of Electronics and Nanoengineering, Aalto University, P.O. Box 15500, FI-00076 Aalto, Finland}

\author{Eleftherios N. Economou}
\affiliation{Department of Physics, University of Crete, 70013 Heraklion, Crete, Greece}
\affiliation{Institute of Electronic Structure and Laser, Foundation for Research and
Technology Hellas, N. Plastira 100, 70013 Heraklion, Crete, Greece}

\author{Maria Kafesaki}
\affiliation{Department of Materials Science and Technology, University of Crete, 70013, Heraklion, Crete, Greece}
\affiliation{Institute of Electronic Structure and Laser, Foundation for Research and
Technology Hellas, N. Plastira 100, 70013 Heraklion, Crete, Greece}

\author{Costas M. Soukoulis}
\affiliation{Ames Laboratory and Department of Physics and Astronomy, Iowa State University, Ames, Iowa 50011, USA}
\affiliation{Institute of Electronic Structure and Laser, Foundation for Research and
Technology Hellas, N. Plastira 100, 70013 Heraklion, Crete, Greece}

\begin{abstract}
Software-defined metasurfaces are  electromagnetically ultra-thin, artificial components that can provide engineered and externally controllable functionalities. The control over these functionalities is enabled by the metasurface tunability, which is implemented by embedded electronic circuits that modify locally the surface resistance and reactance. Integrating controllers within the metasurface cells, able to intercommunicate and adaptively reconfigure it, thus imparting a desired electromagnetic operation, opens the path towards the creation of an artificially intelligent (AI) fabric where each unit cell can have its own sensing, programmable computing, and actuation facilities. In this work we take a crucial step towards bringing the AI metasurface technology to emerging applications, in particular exploring the wireless mm-wave intercell communication capabilities in a software-defined HyperSurface designed for operation in the microwave regime. We examine three different wireless communication channels within the landscape of the reflective metasurface: Firstly, in the layer where the control electronics of the HyperSurface lie, secondly inside a dedicated layer enclosed between two metallic plates, and, thirdly, inside the metasurface itself. For each case we examine the physical implementation of the mm-wave transceiver nodes, we quantify communication channel metrics, and we identify complexity vs. performance trade-offs.
\end{abstract}

\maketitle

\section{Introduction}
\input{1-intro.tex}

\section{Structure, environment description and electromagnetic operations}
\vspace{-0.1cm}
\label{sec:envDes}
\input{2-structure.tex}

\section{Communication in the chip layer}
\label{sec:interchip}
\input{3-chip.tex}

\section{Communication in a dedicated parallel plate waveguide}
\label{sec:dedicated}

\input{4-dedicated.tex}

\section{Communication in the metasurface layer}
\label{sec:dipdip}
\input{5-metasurface.tex}

\section{Conclusion}
\label{sec:conc}
This paper has examined the problem of intercell wireless communication in the complex landscape of \emph{intelligent metasurface fabrics}, where such communication is necessary to implement unique features such as autonomy, interconnectivity, and distributed sensing and intelligence. The introduction of such integrated means of wireless communications is supported by its natural broadcast capabilities, the ease of assembly, and the improved off-chip connectivity. We explored three possible propagation paths at mm-wave frequencies, a band that enables the integration of antennas within HSFs and theoretically avoids interference with the interaction of HSFs with external microwave sources.

Our explorations have clarified the pros and cons of each alternative. The chip layer appears to be a natural choice since the antennas can be integrated on the controller chips and do not interfere with the metasurface operation. Our analysis, however, yields a very large path loss of 40--50~dB at 60~GHz and up to 50--70~dB at 120~GHz, mainly caused by the lossy silicon at the chips. The introduction of a dielectric layer, originally employed for thermal and mechanical support, can decrease the path loss by around 10~dB. The second analyzed alternative consists of the inclusion of a dedicated layer for wireless inter-cell communication. This option yields much lower path loss of 5--25~dB and effective protection to interferences at all bands, but at the expense of (i) a relatively large delay spread in the order of 100~ps for a coherence bandwidth of 10~GHz and (ii) a higher volume and manufacturing cost. Finally, we showed that a promising path loss of 20--40~dB is achievable at 60~GHz without altering the HSF architecture by using the metasurface layer opportunistically as propagation path. This solution, however, maintains the relatively high dispersion of the dedicated layer and requires a careful co-design with the metasurface to minimize interference and signal leakage.

\section*{Acknowledgment}
\vspace{-0.1cm}
This work was supported by the European Union Horizon 2020 research and innovation programme-Future Emerging Topics (FETOPEN) under grant agreement No 736876. Odysseas Tsilipakos acknowledges the financial support of the Stavros Niarchos Foundation within the framework of the project ARCHERS ("Advancing Young Researchers Human Capital in Cutting Edge Technologies in the Preservation of Cultural Heritage and the Tackling of Societal Challenges").

\bibliography{Final}

\end{document}

%% file: 1-intro.tex

Metasurfaces (MSs), the two dimensional version of metamaterials, are planar artificial structures with purposely designed periodically aligned subwavelength features, the unit cells, that provide overall control over the metasurface electromagnetic (EM) properties \cite{Soukoulis:2011,Glybovski2016,chen2016review,Ding2018}. They exhibit a wide variety of exotic electromagnetic functionalities, from perfect and controllable absorption, to beam and wavefront shaping, polarization control, broadband pulse delay, or harmonic generation \cite{Radi2015,Tasolamprou2019720,Wang2018,Perrakis:2019,Tasolamprou201423147,Pfeiffer20153248,Asadchy2017,Tsilipakos2018,Fu:2019,Gansel20091513,Tsilipakos:2018acs,Sounas2013,Li2015607}. 
To add reconfigurability, together with the ability to host multiple functionalities within a single MS, recent works have proposed to embed circuits capable of tuning the response of each individual unit cell \cite{Makarov2017,Liu2018,Shaltout2019}. This allows us   to drive such reconfigurable MSs via a centralized control unit such as a field-programmable gate array (FPGA) \cite{Cui2014,Yang:2016}. With the addition of external sensors providing feedback to the control unit, adaptive MSs can be conceived.

The recent HyperSurFace (HSF) paradigm aims to make a firm step ahead by realizing the vision of an \emph{intelligent metasurface fabric} where each unit cell not only incorporates its own sensing, programmable computing and actuation facilities, but can also exchange information with other unit cells \cite{Liaskos2015,Pitilakis2018MMParadigm}. This allows the MS (i) to be autonomous, adapting to the environment without external intervention, (ii) to be seamlessly interconnected with other MSs, and subsequently (iii) to implement distributed sensing and intelligence both at the MS or system levels. This approach enables many exciting applications in robotics or within the exploding spectrum of ultra-compact Internet-of-Things (IoT) platforms such as high-capacity wireless networks, physical-layer security, intelligent wireless environments, distributed beamforming and spatial-index modulations \cite{Liaskos20191,Liaskos2019JointCS,Renzo2019,Basar2019,hu2018beyond}.

The integration of the computing and intercell communication circuits on a per-cell basis is critical for the realization of the HSF vision \cite{AbadalACCESS,Tasolamprou2018, Petrou2018}. A first implementation, as we will see, considers the embedding of multiple chips within the MS structure. Thus, by definition, we need to interconnect (i) the controllers within the same chip and (ii) the multitude of chips within the HSF to provide the highly sought distributed intelligence. In such an integrated environment, wired communication is the default choice as technical know-how from general-purpose Network-on-Chip (NoC) or low-power embedded systems can be effectively used \cite{Bjerregaard2006, Bertozzi2014}. However, for off-chip communication, input/output pin scarcity severely limits the bandwidth and connectivity, whereas the latency and power consumption of transmission links increases significantly with distance. Even within the chip, NoCs may still suffer from power and latency issues in very dense HSFs due to their resemblance to massive manycore processors, where such issues are well known \cite{Grot2011}. In such cases, wireless technology has proven to reduce the broadcast latency by an order of magnitude with a similar energy consumption \cite{Fernando2019}. 

\begin{figure}[!t]
\vspace{-0.2cm}
\centering
\includegraphics[width=\columnwidth]{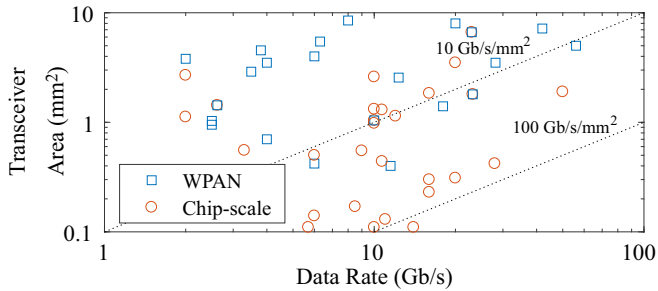}
\vspace{-0.2cm}
\caption{Area as a function of the data rate for state-of-the-art transceivers for Wireless Personal Area Networks (WPAN) and chip-scale applications. Data extracted from \cite{Tasolamprou2018} and references therein.}\label{figIntro} 
\vspace{-0.2cm}
\end{figure}


In light of these drawbacks, wireless intercell communication becomes a compelling alternative in large or dense HSFs. The wireless option has the obvious advantage of not requiring wiring between the chips, which facilitates the assembly and improves the modularity of the solution. Moreover, assuming omnidirectionality, the wireless technology naturally supports broadcasting, thereby facilitating data dissemination and the implementation of distributed intelligence. Such enhanced connectivity also allows to implement denser inter-chip network topologies. These advantages are analogous to those in on-chip networks \cite{Matolak2012,Fernando2019} and other computing systems \cite{Shin2012}. We note that wireless intercell communication refers to information transfer among the unit cells of a single HSF device and that wireless signals are not intended to be leaked outside the HSF.

The wireless approach is enabled by recent advances in on-chip antennas in mm-wave and THz bands \cite{Markish2015, Gutierrez2009, Cheema2013}, as well as the constant miniaturization of RF transceivers for short-range applications. As shown in Fig.~\ref{figIntro}, transceivers with multi-Gbps speeds and footprints as small as 0.1 mm\textsuperscript{2} have been demonstrated. Before assessing the potential applicability of existing transceivers, though, a deep understanding of the electromagnetic wave propagation within this new landscape is essential. For instance, it is important to analyze the possible propagation paths and assess the potential interference between the MS operation and the wireless network. Some researchers have studied propagation in environments with metallic enclosures similar to HSFs \cite{Genender2010, Ohira2011, Khademi2015}, whereas others have investigated propagation within computing packages \cite{Branch2005, Zhang2007, Kim2016mother}. However, these works consider structures that differ considerably from HSFs and do not account for their particularities in terms of RF interference.

In this paper, we undertake the thorough analysis of wireless mm-wave intercell communication in the HSF environment. The wireless intercommunication of the metasurface cells is a major step towards the realization of  artificially intelligent fabrics. It provides the HyperSurface the capability of interchanging information through the whole extend of the fabric. This allows the controllers to readapt the metasurface according to the external conditions even if modified locally, compensate for any malfunctioning cells, select configurations that balance power consumption and many others functionalities assessed with the readily assembled wireless solution. In \cite{Tasolamprou2018}, we performed a preliminary assessment of the spectral characteristics of intercell communication in two specific channels and in the frequency domain.
Here, we present a rigorous frequency and time domain study towards the exploration of the wireless communication channels within the landscape of a fully functional HSF, consisting of the components effectuating the electromagnetic manipulation of microwave radiation, the MS layer and the controller chip.
We consider three different communication channels with different advantages and constraints. In Section \ref{sec:interchip}, we assess the chip area as a communication channel, a natural choice since there lie all the electronics of the device. Then, in Section \ref{sec:dedicated}, we assume a parallel plate waveguide dedicated solely to the intercell mm-wave communication, a choice of high efficiency and security, at the expense of device complexity and volume. Finally, in Section \ref{sec:dipdip}, we examine the metasurface layer itself as an opportunistic intercell communication channel which minimizes the volume of the device. In all channels, we evaluate the communication performance through frequency- and time-domain numerical calculations providing data for the channel transfer function,  mean delay and delay spread. Technical considerations, detailed designs, analysis of obtained simulation results, and discussion are provided for each channel, separately, before drawing the overarching conclusions and outlook of the intercell communication exploration, compared to our preliminary study \cite{Tasolamprou2018}, in Section \ref{sec:conc}.


%% file: 2-structure.tex
\subsection{Environment Description}

\begin{figure*}[ht]
\centering
\includegraphics[width=\textwidth]{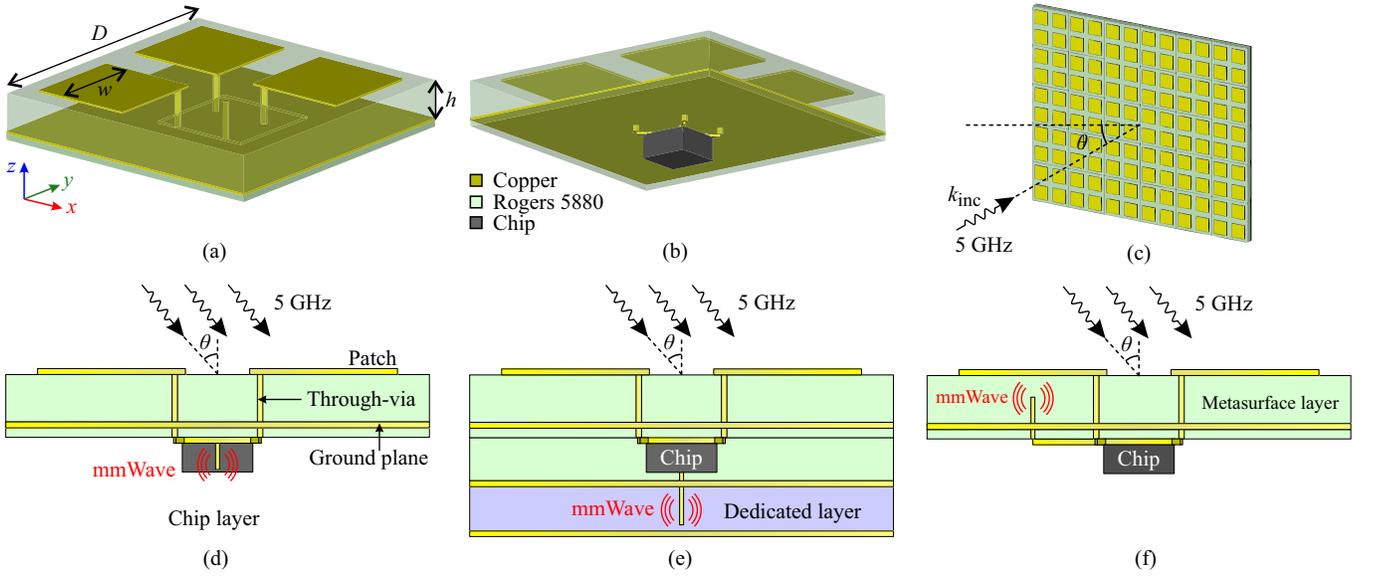}
\vspace{-0.2cm}
\caption{The HyperSurface under study. (a) Top-view of unit cell with geometric parameters and (b) bottom-view of unit cell with the controller chip for the programmable operation. (c) A $5\times6$ unit cell MS operating at 5 GHz under oblique plane wave incidence. (d,e,f) Unit cell side-views (vertical cross-sections) illustrating the three considered communication channels: (d) Communication in the chip layer, (e) communication in a dedicated parallel-plate waveguide and (f) communication inside the metasurface dielectric.}\label{fig1}
\vspace{-0.2cm}
\end{figure*}

As a case study we consider the software-defined HSF depicted in Fig.~\ref{fig1}. The metasurface (MS) part shown in Fig.~\ref{fig1}a consists of a periodic array of electromagnetically thin metallic patches placed over a dielectric substrate  back-plated by a metallic layer serving as ground. To enable the software-based MS control, the patches are connected to a group of controller chips that lie below the metallic back plane through vertical vias (isolated from ground by holes) as shown in Fig~\ref{fig1}a. The characteristics of the impinging electromagnetic wave that is reflected from the metasurface (see Fig.~\ref{fig1}c) can be controlled depending on the electromagnetic features of each unit cell.  The controllers adjust the electromagnetic behaviour of the metasurface fabric by attributing additional local resistance and reactance at will \cite{Cui2014,Yang:2016,Fu:2019}. The controller plane is decoupled from the MS thanks to the back plane that separates the patches from the chips.

Our case study MS is designed for perfect tunable absorption and anomalous reflection operation in the microwave regime. For operation in the microwave regime, the lateral size of the metasurface is required to be in the order of tens of centimetres, i.e., at least a few wavelengths. Specifically, the reference MS structure under consideration is designed to operate at $f=$ 5 GHz ($\lambda_0 \sim 60$~mm). It consists of periodically arranged, four-patch unit cells with $xy$ size equal to $D~\times~D~=~$~12~mm ~$\times$~12~mm, as shown in Fig.~\ref{fig1}. The size of each patch is $w~\times~w~= $~4.2~mm~$\times$~4.2~mm. The thickness of the substrate is $h=1.575$ mm and it is made of Rogers RT/Duroid 5880 with permittivity $\epsilon_r=2.2$ and loss tangent $\tan\delta=9\times10^{-4}$.	 Another high frequency laminate board could be also appropriate for this work.

\begin{figure}[ht]
\centering
\includegraphics[width=0.9\columnwidth]{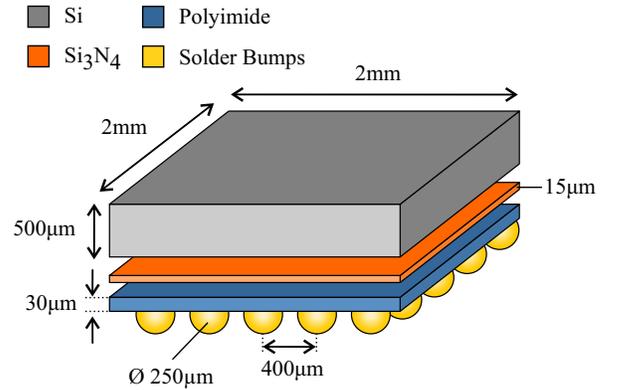}
\vspace{-0.2cm}
\caption{\label{fig:chip}  Layout of the chip hosting the metasurface controllers. The stackup includes a low-conductivity silicon-die, a silicon nitride insulation layer, and a polyimide passivation layer. The total volume of the chip is approximately 2 mm $\times$ 2 mm $\times$ 0.8~mm.}
\vspace{-0.2cm}
\end{figure}

We assume at this point  that each chip serves four metallic patches. Chips are square with lateral dimensions of 2~mm X 2~mm and are placed 0.15 mm below the back plane. In our case study, we consider a conventional flip chip package which mainly consists of a standard die with attached solder bumps that carry the input/output (I/O) signals.  The flip chip package will be assembled with bumps facing down on the backplane. Figure \ref{fig:chip} shows a chip schematic layout, which includes, from top to bottom, a stack of materials typically found in conventional chips: low-conductivity  silicon die ($\epsilon_r=11.9$, $\rho = 10~\Omega$-m, thickness of 0.5 mm), an insulator (silicon nitride of thickness 15 $\mu$m, $\epsilon_r=7.5$, assumed lossless), and a polyimide layer ($\epsilon_r=3.5$, 30-$\mu$m thick) that acts as passivation. The passivation layer allows to have an extra metallic layer, which in this case is a copper redistribution layer that connects the first metallization layers of the chip with the corresponding solder bumps at the bottom. In our case study, the solder bumps have a diameter of 0.25 mm and a pitch of 0.4 mm. We consider that the space surrounding the solder bumps, between the flip chip and the board, is free of underfilling epoxy material usually added for mechanical robustness. We further assume that the metasurface is not mounted on top of any object and, therefore, that there is nothing preventing signals to propagate below the chips.

\subsection{Wireless Propagation Paths}
The physical landscape of the software-defined HSF offers several opportunities for the propagation of RF signals within the structure for wireless connectivity between the different controllers. The actual implementation depends on the tile lateral dimensions and the targeted wavelength. In this work we consider three distinct communication channels, as seen in Fig.~\ref{fig1}(d,e,f):

\begin{itemize}

\item \textbf{Communication in the chip layer}. The first channel under consideration is the back side of the metasurface, where the chip and additional circuitry lie, together with any system packaging; it should be noted no structural change to the HSF landscape is assumed. Monopoles can be implemented by means of TSV within the chips or regular vias placed on the chip sides. Information propagates omnidirectionally through the system package and part of it penetrates into the chips. Wireless communication in the chip layer is a natural choice since it is the section where all the electronics of the software defined metasurface are placed. The evaluation of this communication channel is presented in  Section \ref{sec:interchip}.

\item \textbf{Communication in a dedicated layer} The second channel under investigation is a dedicated communication layer formed by adding two extra metallic plates below the chip. Monopoles fed from the chip are inserted in the parallel-plate waveguide formed, and excite waves that propagate omnidirectionally in this environment, the only obstacles being the monopole antennas. This communication channel is completely isolated from the core of the metasurface and therefore does not interfere with the metasurface operation which offers increased security.  Section \ref{sec:dedicated} evaluates the dedicated layer communication.

\item \textbf{Communication in the metasurface layer} The last channel under consideration is the space between the metasurface patches and the metallic ground plane, called MS layer. A blind via fed from the chip serves as the antenna, while the metallic patches and the metallic back plane act as a waveguide. The communication and metasurface operation takes place in the same volume which minimizes the overall size of the device. Section \ref{sec:dipdip} evaluates the communication scenario in the metasurface layer.

\end{itemize}

We study the different  channels with respect to their physical constraints and the communication opportunities that they offer. In both  chip and  dedicated layer communication channels,  the wave propagates in a restricted waveguide. 
The fact that they are completely decoupled from the metasurface layer allows for a wide choice of communication frequencies since the corresponding parts of the structure can be adapted accordingly without affecting the operation of the metasurface. On the other hand, the  communication channel in the metasurface layer needs to be designed respecting the metasurface operation which requires a specific and invariable geometry. This posses constraints in the design of the communication channel which will affect the details of the communication.  

\subsection{Operational Conditions}
To ensure that the electromagnetic response of the metasurface and the wireless communication operation are decoupled for all communication paths, we choose the communication frequency to be greater than 25 GHz. This decoupling is especially important in  the case where the metasurface layer hosts both the electromagnetic waves  for the metasurface operation as well as the communication signals which is seen in Fig.~\ref{fig1}(f) and analysed in Section \ref{sec:dipdip}. Therefore, in overall, we investigate the channel communication in the range $f=$~25 to 200 GHz, although we could investigate lower frequencies for the chip and dedicated channels. The distance between two neighbouring nodes equals $D=12$~mm which, in the frequency regime under consideration, is in the order of 5$\lambda_0$ to 40$\lambda_0$, respectively. This means that the communication takes place in the near and intermediate field regime. Thus, we cannot resort to simplified farfield calculations and we use full wave electromagnetic analysis for the numerical investigation. For higher frequencies, i.e., for frequencies $f>1$ THz ($D>200\lambda$) the full wave analysis becomes cumbersome and we would need to turn to simplified schemes such as ray tracing \cite{Kantelis20141823}. It is stressed that even though we perform the analysis for the reference case dimensions, a direct scaling of the structure along with the wavelengths of operation is possible assuming that the properties of the materials involved remain the same.


\subsection{Performance Metrics}
In the simulations we generally consider a number of antennas evenly distributed across the simulated space. The outcome is the field distribution, the antenna gain, and the coupling between antennas. To see the electromagnetic coupling (communication) between any pair of antennas, it is enough to observe the magnitude of the scattering matrix component $S_{21}$ (a complex number). Generally speaking, for a two-port network the scattering matrix is described as
\begin{equation}
S=\begin{bmatrix}
S_{11}&S_{12}\\
S_{21}&S_{22}
\end{bmatrix},
\end{equation}
where $S_{21}$, $S_{12}$ determine the transmitted power from one port to the other, and $S_{11}$, $S_{22}$ represent the power reflected out of the network from each port, owing to impedance mismatch. Due to the reciprocity and symmetry of the problem, it holds that $S_{12}=S_{21}$ and $S_{22}=S_{11}$, respectively.  In the structures under study, however, in addition to $S_{21}$, we also observe the $S_{11}$ component in order to understand how much of the power is reflected back to the antennas operating as the transmitter/receiver. In the optimum case the antennas need to be impedance matched, that is, we need  to achieve a very low value for the magnitude of $S_{11}$. This translates into negligible losses due to  reflection, and  high values for the magnitude of $S_{21}$. However, in the case of noticeable return loss (reflection), we can resolve this issue by employing an external matching circuit. In fact, the important parameter is essentially the transmission coefficient. We note here that the formulation above can be generalized for any transmitter $i$ and receiver $j$.

\textbf{Frequency Domain Analysis.} Besides the $S$-parameters, a metric pertinent to channel characterization is the transfer function, $\vert H_{ij}(f)\vert$, of the channel defined between the transmitter $i$ and the receiver $j$. It is obtained via the following expression
\begin{equation} \label{eq:Hf}
G_{i} G_{j} \vert H_{ij}(f)\vert^{2} = \frac{|S_{ji}(f)|^{2}}{(1 - |S_{ii}(f)|^{2})\cdot(1 - |S_{jj}(f)|^{2})},	
\end{equation}
where $G_{i}$ and $G_{j}$ are the transmitter and receiver antenna gains, while $S_{ji}$, $S_{ii}$, and $S_{jj}$ are the scattering matrix elements defined above. Once the whole matrix of frequency responses $\mathbf{H}$ is obtained, a path loss analysis can be performed by fitting the attenuation $L$ over distance $d$ to
\begin{equation} \label{eq:PL}
L = 10\,n \cdot \log_{10}(d/d_{0}) + L_{0},
\end{equation}
where $L_{0}$ is the path loss at the reference distance $d_{0}$ and $n$ is the path loss exponent \cite{Zhang2007}. The path loss exponent is around 2 in free space, below 2 in guided or enclosed structures, and above 2 in lossy environments. Since losses in the channel are crucial to determine the power consumption at the transceiver we will report improvements in terms of worst-case $L_{\text{max}}$, average $L_{\text{avg}}$, and path loss exponent $n$. In the cases where the antennas are matched and their gain is small, $S_{ji}$ alone provides a fair indication of path loss.


\textbf{Time Domain Analysis.} In the time domain, the electromagnetic (EM) solver allows to define an input excitation $x(t)$ at the  transmitting antenna. The impulse response $h_{ij}(t)$ between transmitter $i$ and receiver $j$ can be derived by the input transmitting signal and the output signal $y(t)$ obtained at the antennas using the classical formulation:
\begin{equation} \label{eq:ht}
y_{j}(t) = x_{i}(t) \star h_{ij}(t),
\end{equation}
where $\star$ denotes the convolution operator. Once the impulse response is calculated, it is straightforward to evaluate the PDP as
\begin{equation} \label{eq:PDP}
P_{ij}(\tau) = |h_{ij}(t,\tau)|^{2},
\end{equation}
therefore obtaining a matrix of PDP functions $\mathbf{P}$. To characterize the channel in the time domain, we first obtain the \emph{mean delay} $\overline{\tau_{ij}}$ defined as
\begin{equation}
\label{eq:meanDelay}
\overline{\tau_{ij}} = \frac{\int{\tau P_{ij}(\tau) d\tau}}{\int{P_{ij}(\tau)\, d\tau}}.
\end{equation}
We also evaluate the multipath richness of the channel by obtaining the delay spread $\tau_{\text{rms}}$ of each PDP as 
\begin{equation}
\label{eq:DelaySpread}
\tau_{\text{rms}}^{(i,j)} = \sqrt{\frac{\int{(\tau - \overline{\tau_{ij}})^{2} P_{ij}(\tau)\, d\tau}}{\int{P_{ij}(\tau)\, d\tau}}},
\end{equation}

From a communications channel perspective, the delay spread provides a lower bound of the signal bandwidth that can be decoded correctly. Such measure is generally referred to as coherence bandwidth and can be calculated for a channel between transmitter $i$ and receiver $j$ as 
\begin{equation}
\label{eq:Bc}
B_{c}^{(i,j)} = \frac{1}{\tau_{\text{rms}}^{(i,j)}}.
\end{equation}

In this work we will assume that all wireless channels are broadcast and, therefore, they should be operated at the lowest speed ensuring correct decoding at all nodes. As a result, we will take the worst delay spread as limiting case and use it to evaluate the coherence bandwidth $B_{c}$, as follows
\begin{equation}
\label{eq:Bc2}
\tau_{\text{rms}} = \max_{i,j\neq i}{\tau_{rms}^{(i,j)}} \Rightarrow B_{c} = \frac{1}{\tau_{\text{rms}}}.
\end{equation}

%% file: 3-chip.tex
In this scenario, antennas are integrated within the chip or at its close vicinities and propagation occurs in two regions: (i) the intra-chip region, in which the waves radiated by the monopole inside the silicon substrate travel through several layers of the chip (mainly the silicon layer); and (ii) the inter-chip region, in which the waves that have left the chip travel through the inter-chip space until they reach the boundaries of another chip.

The main advantage of this configuration is that the communication channel does not require significant modifications with respect to  the original structure. This leads to a very cost-effective implementation that leaves the antenna and transceiver circuits as the only elements that may incur some area overhead. In fact, antenna and transceiver integration are performed for each chip individually, decoupling this process from the complete system integration and thereby reducing its complexity.

Depending on the actual implementation of the system package and on the mounting of the HSF within the propagation environment (e.g., over a wall), this scenario could lead to a totally enclosed volume. This would exclude the possibility of any coupling between `external' electromagnetic radiation (e.g., metasurface operation microwaves or other stray incoming signals) and the intercell communication. Still, losses may still arise due to reflections, dissipation in the chip materials, or spreading in undesired areas within the package. In this work, however, we will not make specific assumptions on system packaging or the mounting of the HSF, leaving the space among and below the chips empty (i.e. filled with air).

\subsection{Simulation results}
We first simulate a scenario with 5$\times$5 chips embedded within the HSF. We model the antenna as a TSV at the center of each chip and adjust the length to build antennas with fundamental resonance at $f_{1} = 60$ GHz, $f_{2} = 90$ GHz, and $f_{3} = 120$ GHz. To this end, the lower layers of the chip (the redistribution layer and the array of solder bumps) act as an effective ground plane. Therefore, the TSV can be modeled as a quarter-wave monopole. Since neighboring chips are at a distance of 12 mm, the relative distance among adjacent antennas in terms of the free-space wavelength ranges from 2.4$\lambda_0$ (60 GHz) to 4.8$\lambda_0$ (120 GHz). To perform the analysis, we excite the antenna in the bottom-left corner of the system.

\begin{figure}[!t]
\centering
\includegraphics{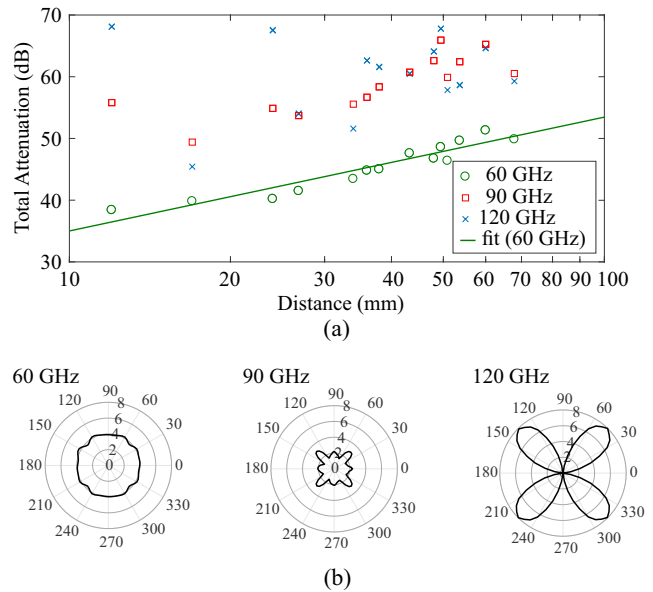}
\vspace{-0.2cm}
\caption{\label{fig:Fig4}Frequency domain analysis of the chip-layer communication scenario for three different frequencies: 60, 90, and 120 GHz. (a) Total attenuation as a function of distance and (b) directivity patterns.}
\vspace{-0.2cm}
\end{figure}

Figure \ref{fig:Fig4}(a) shows the path loss as a function of the distance. The first aspect worth noting is the rather large attenuation in the range of 40--70 dB, which is in part due to the low gain of the antennas placed within the lossy silicon. It is also observed that the losses increase with the distance, as expected. Since the inter-chip medium is air, we can attribute these increasing losses to the reflections caused at the silicon-air interfaces and the spreading of energy in the half-space below the chips. Another remarkable result is the significantly larger attenuation observed at $f_{2} = 90$ GHz and $f_{3} = 120$ GHz. This is caused by both the larger spreading losses at such frequencies and the low directivity of the antennas at certain directions. The second effect is justified in Fig. \ref{fig:Fig4}(b): the use of a monopole within a squared cavity of fixed size leads to a pattern that may be omnidirectional or favor certain directions depending on the communication frequency. We can thus conclude that on-chip antenna design has a great impact on the path loss and should be carefully approached when implementing inter-cell communication within the chip layer. 

\begin{figure}[!t]
\centering
\includegraphics{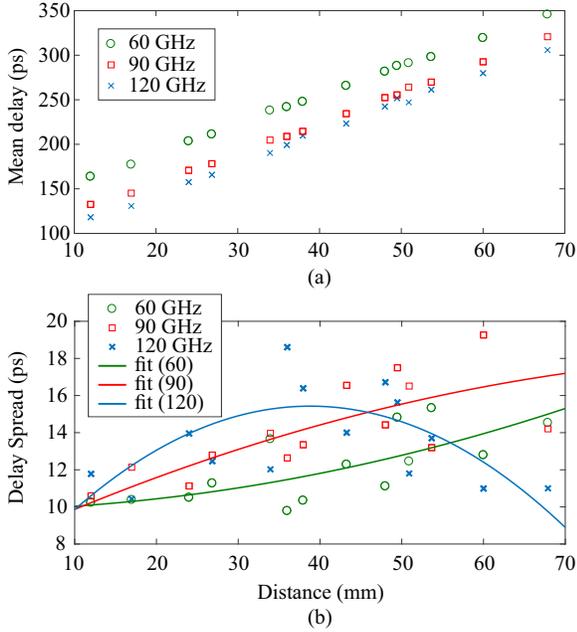}
\vspace{-0.2cm}
\caption{\label{fig:Fig5}Time domain analysis of the chip-layer communication scenario for three different frequencies: 60, 90, and 120 GHz. (a) Mean delay and (b) delay spread.}
\vspace{-0.2cm}
\end{figure} 

In the time domain, we obtain the impulse response and calculate the mean delay and delay spread in the system. Figure \ref{fig:Fig5} plots both metrics in each chip in a 5$\times$5 array as a function of the distance. We first note that the mean delay follows a linear dependence with the distance, which suggests that a significant part of the energy is transported by the line-of-sight ray. The communication frequency is of minor relevance here. As for the delay spread, we cannot observe clear scaling trends with frequency or distance. In general, delay spread would increase with distance because the main ray has lower energy and reflections become significant. This behavior can be somehow inferred for $f = 60$~GHz. At higher frequencies, however, the behavior tends to differ due to: (i) the directional radiation patterns and (ii) the fact that chips located in the center of the system are completely surrounded by other chips that could increase the number of relevant reflections. This justifies the higher delay spread at distances around 30--50~mm. In any case, the worst case delay is lower than 19.26~ps, which yields a coherence bandwidth over 51.92~GHz. 


\subsection{Enhancing inter-chip propagation via surface waves}
The results shown above confirm that losses, and non-multipath effects, would be the main impairment of communication in the chip layer. One of the reasons is that the system is not completely enclosed and waves may spread out away from the chips. Within this context, it would be interesting to create a wireless channel that propagates around the chips following a path along the surface of the MS back plane. Surface waves are bound states that propagate  along the interface of two semi-infinite domains and exhibit many interesting features \cite{Tasolamprou201513972,Tasolamprou20172782a,Kwon2018,Fan2019169}.
A dielectric material close to a metallic plane can support a TM surface wave that travels along the metal-dielectric interface. For this to work, the dielectric layer needs to have a sufficiently high reactance \cite{OpokuAgyeman2016}. The reactance $X_{s}$ is calculated as
\begin{equation} \label{eq:reactance}
X_{s} = 2 \pi f \mu_{0} \bigg[ \bigg(\frac{\epsilon_{r}-1}{\epsilon_{r}}\bigg)t + \frac{1}{2}\Delta\bigg]
\end{equation}
where $\varepsilon_{r}$ is the permittivity of the material, $t$ is the layer thickness, and $\Delta$ is the skin depth of the conductor at the frequency of operation, which is given by $\Delta = (\pi f \mu_{0} \sigma)^{-1/2}$. The term $\sigma$ refers to the conductivity of the metallic plane. To play this role, we propose to add a layer of Aluminium Nitride (AIN, $\epsilon_r=8.6$, lossless) as a common interface between all the chips and the back plane. This material is typically used as a heat spreader in chip packages \cite{Timoneda2018b} and would be compatible with the MS fabrication processes. We thus propose to fill the space between the MS back-plane and the chips (0.15 mm in this paper) with the AIN material. 

\begin{figure}[!t]
\centering
\includegraphics{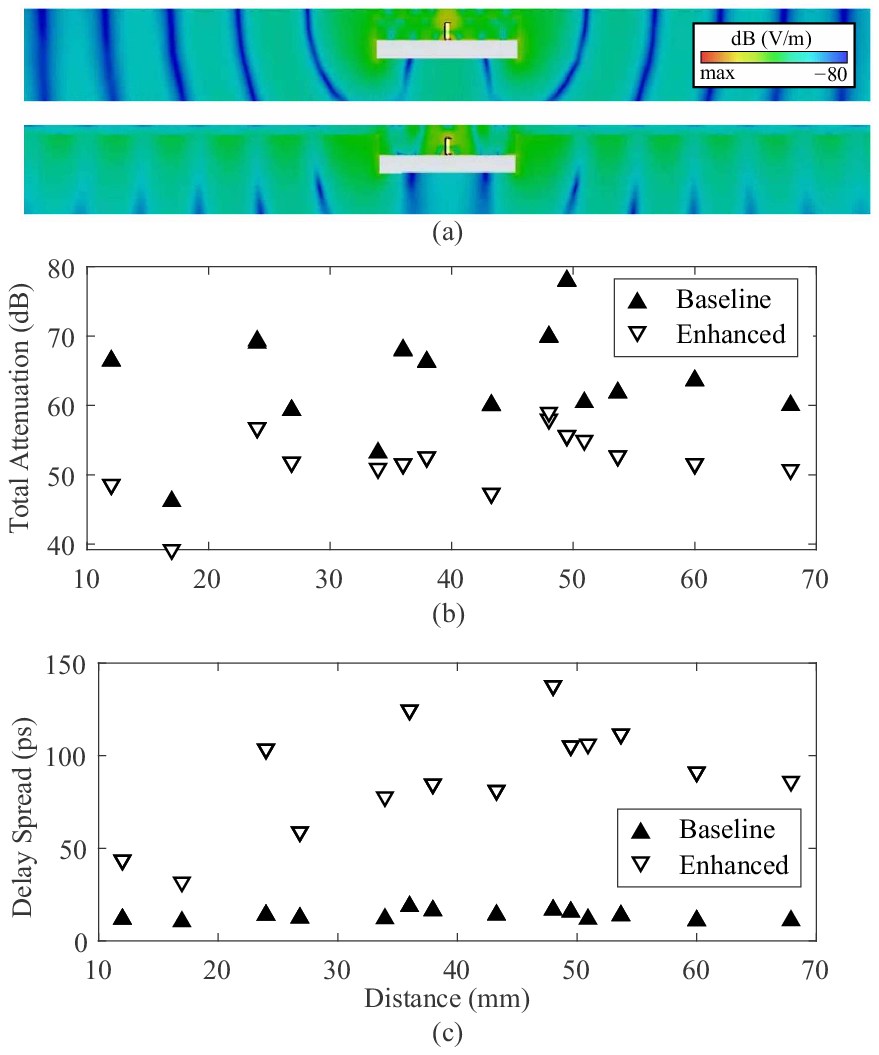}
\vspace{-0.2cm}
\caption{\label{fig:Fig6}Effects of adding a dielectric layer to enhance propagation for the case $f = 120$ GHz. (a) Electric field distribution at the surroundings of the radiating chip without and with dielectric layer (top and bottom, respectively). (b) Path loss comparison. (c) Delay spread comparison.}
\vspace{-0.2cm}
\end{figure}

Figure \ref{fig:Fig6}(a) shows the electric field distribution at the inter-chip space at 120 GHz. The top chart illustrates how, as expected, most of the power radiates away from the metallic back plane without the dielectric layer. The bottom chart, on the other hand, demonstrates that the dielectric layer is able to bind the surface waves and reinforce propagation along the dielectric in the path between chips. 

To evaluate the improvement in terms of path loss, we obtained the channel attenuation with and without the  dielectric layer for $f = 120$~GHz. As it is shown in Fig. \ref{fig:Fig6}(b), the path loss is reduced significantly. In average, the improvement is of 12.24~dB with maximum values around 30~dB, demonstrating that the dielectric layer can be an enabler of wireless communication in the chip layer. This reduction in losses, however, comes at the cost of a significant increase in the delay spread due to the leaky waveguide effect of the dielectric layer. To verify this, we repeated the time domain analysis with and without the dielectric layer and compared the results. The mean delay, not shown for brevity, increases significantly for the dielectric case as now signals propagate through a material with higher permittivity. As for the delay spread, Fig. \ref{fig:Fig6}(c) shows how the value for the dielectric-enhanced case rises by almost an order of magnitude with respect to the baseline. The worst-case delay spread reaches a value of 137.74~ps, leading to a coherence bandwidth of 7.24~GHz. At the other frequencies, we observed similar behaviors with different relative values. For instance, the path loss improvement is more modest at $f = 60$~GHz, around 6~dB in average, since the coupling to the dielectric layer is diminished due to its lower reactance.

\subsection{Discussion}
The results obtained in this section lead to several conclusions. First, the use of the chip layer for inter-cell wireless communication decouples the problem of antenna integration from the system perspective, but leaves it to the chip designers instead. The choice of the communication frequency needs to take into account the antenna placement and dimensions of the chip: the silicon die may act as a resonant cavity and modify the radiation pattern of the antenna, impacting on the expected path loss. Furthermore, the thickness of the silicon layer imposes a lower bound on the frequency of the communication as it determines the maximum practical length of the resonant monopole. Therefore, system architects may need to balance out performance gains arising for the use of lower frequencies (larger antenna apertures with lower spreading losses) against the potential cost implications of increasing the size of the chips. Finally, it seems clear that the traditional tradeoff between path loss and delay spread is also apparent in this scenario: the addition of a dielectric layer improves the former but worsens the latter. It is in fact provable that the thickness of the dielectric can be used to tailor this tradeoff as, theoretically, thinner layers would lead to better delay spread (and worse path loss) figures while thicker layers would yield opposite results.

%% file: 4-dedicated.tex
Another opportunity for the intercell communication channel is the isolated layer depicted in Fig. \ref{fig1}(e). The channel is dedicated exclusively  to transferring the signals between the communication nodes. It is implemented by introducing two additional metallic plates below the chip which form a parallel-plate metallic waveguide channel whose thickness, as explained later on, is specified by the desired frequency of operation.  The space between the two metallic plates is empty or filled with a uniform dielectric material; here we assume that it is empty (air).  In each chip we connect a probe antenna which extends in the parallel-plate waveguide space through a small perforation in the metallic plate  below the chip, as shown in Fig. \ref{fig1}(e). The length of the wire antenna is assumed to be approximately equal to the height of the parallel-plate waveguide (actually a small gap to avoid short circuit is assumed) and it radiates omnidirectionally, i.e., a device that transmits or receives electromagnetic power isotropically in the horizontal $xy$-plane. The main advantage of the present configuration is that the communication channel is electromagnetically isolated from the rest of the system, that is, no electromagnetic coupling between the processor and metasurface operation and the communication is expected. Moreover, the parallel plates create a closed space where no energy leakage is allowed (the holes are electromagnetically small) and thus communication security is ensured beyond any doubt. Additionally, there are no obstacles in the propagation space, apart from the probe antennas themselves. For these reasons, this option offers robustness and design flexibility, although it requires  additional fabrication effort and increases the overall volume of the unit cell.

\begin{figure}[h]\centering
\includegraphics[width=88mm]{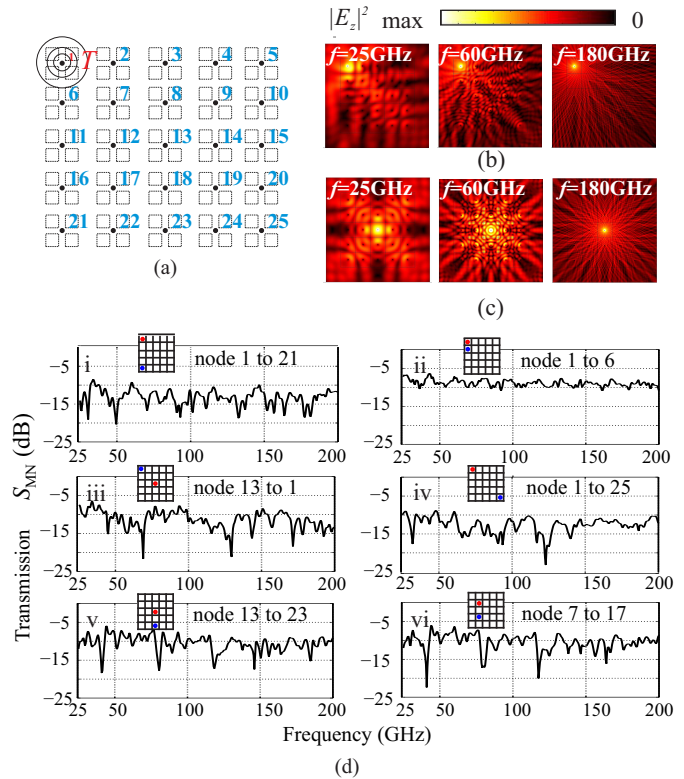}
\vspace{-0.2cm}
\caption{\label{fig:fig_iso_pairs} (a) Schematic of the TEM parallel waveguide 2D approximation, probe no.1 radiates. (b) and (c) Electric field intensity distribution at $f=$~25, 60 and 180~GHz when the emitter is (b) no.~1, in the corner, and (c) no.~13, in the middle, respectively. (d) Power fraction received at the node $M$ when node $N$ radiates, $S_{\text{MN}}$, over the frequency range $f=$~25 to 200~GHz. Six cases of $MN$ node pairs are schematically depicted in the insets. Red/blue is for transmitter/receiver. }
\vspace{-0.2cm}
\end{figure}

The dedicated channel is custom-built to optimize wireless propagation. The metallic plates perfectly reflect the electromagnetic waves and they can guide a discrete spectrum of Transverse Electric (TE) and Transverse Magnetic (TM) modes depending on the distance of the plates $d$, the dielectric permittivity of the filling materials $\varepsilon_r$, and the frequency of operation $f$. A distinctive feature of the  parallel-plate waveguide is that it sustains the propagation of TEM (Transverse Electromagnetic) waves in which both the electric and magnetic fields are perpendicular to the propagation direction. The TEM mode can be excited from zero frequency (DC) and is the only propagation mode supported by the waveguide up to the cut-off frequency of the first higher-order mode: $f<{c_0}/({2d}\sqrt{\varepsilon_r})$. In our implementation we purposely design the dimension so that the frequency lies below the cut off and therefore single mode operation is forced. Since the EM energy is carried by the single TEM mode, the waveguide is naturally impedance matched with free-space; this allows the following approximation: We consider that the propagation in the 3D waveguide can be approximated by a 2D analogue where the monopoles are replaced by finite-size conducting scatterers, placed at the vertical positions of the antenna probes. Each scatterer radiates 2D cylindrical waves in the surrounding space and diffracts the energy coming from the environment. The field radiated from the emitter and the diffracted field from the scatterers interfere creating destructive or constructive patterns in the waveguide.

\subsection{Simulation results}

The 2D approximation allows us to solve for large areas and frequency spans  which provides us with a qualitative evaluation of the propagation properties in a multiscattering environment. A priori, we assume that the antennas are impedance matched in all the spectrum of interest and that only the TEM mode is excited, both effectively controlled by the height of the structure. We investigate the system of 5$\times$5 nodes depicted in Fig. \ref{fig:fig_iso_pairs}(a). Each antenna (scatterer) is a finite size copper cylinder of radius $R = 0.12 $~mm. The distance  between two neighbouring antennas, that is the pitch of the antenna grid, is equal to $D=12$~mm, which is required for the metasurface operation at 5~GHz. 
In the 2D approximation we do not take into account the impedance characteristics of the antennas. The emitter is simulated as a field source that radiates omnidirectional electromagnetic waves. All the surrounding scatterers reflect the incoming wave. In this way we estimate the electric field intensity profile of the propagating waves in the presence of the reflecting obstacles. Figures \ref{fig:fig_iso_pairs}(b,c) present the profile of the total radiated intensity at frequency $f=$~25, 60 and 180~GHz when the emitter is no.~1 and no.~13, respectively. The electromagnetic waves interfere either destructively or constructively producing patterns of high or low intensity corresponding to the bright and dark spots.  Moreover, due to the symmetry of the configuration, the system presents a four fold symmetry, as evidently shown in the figure. This means that many communication pairs and communication paths have by definitions identical properties.

As the communication quality depends both on the position of the antenna probes and the frequency, it is useful to evaluate the connection between separate nodes. This is achieved by estimating, in the position of the receiver, the power that can be captured by the multipath propagation coming from all directions. The total power accumulated in the position of the receiver $M$ when $N$ emits, $P_{\text{MN}}$, is normalized by the total radiated power from the emitter $P_0$. The system is reciprocal, that is, $S_{\text{MN}} = S_{\text{NM}}$. Figure \ref{fig:fig_iso_pairs}(d) presents the power fraction received in the position $M$ transmitted from emitter $N$ over the frequency range of $f=$~25 to 200~GHz for node  pairs  schematically depicted in the insets. As observed in all cases, the received power remains on average the same for each pair in the entire frequency span. However, for nearly all cases, there are some frequency points where the received power drops. For example, for the case of the pair no.~7-no.~17 (panel vi) there appear three dips in the received power at around $f=$~45, 80 and 115~GHz. These points correspond to destructive wave interference. Moreover, we can observe the general tendency of the decreased received power with respect to the node-pair distance, i.e., for the pair no.~1-no.~21 (panel i) the average received power is $-15$~dB whereas for the pair no.1-no.6 (panel ii), the received power is on average $-8$~dB. 
\begin{figure}[ht]\centering
\includegraphics[width=88mm]{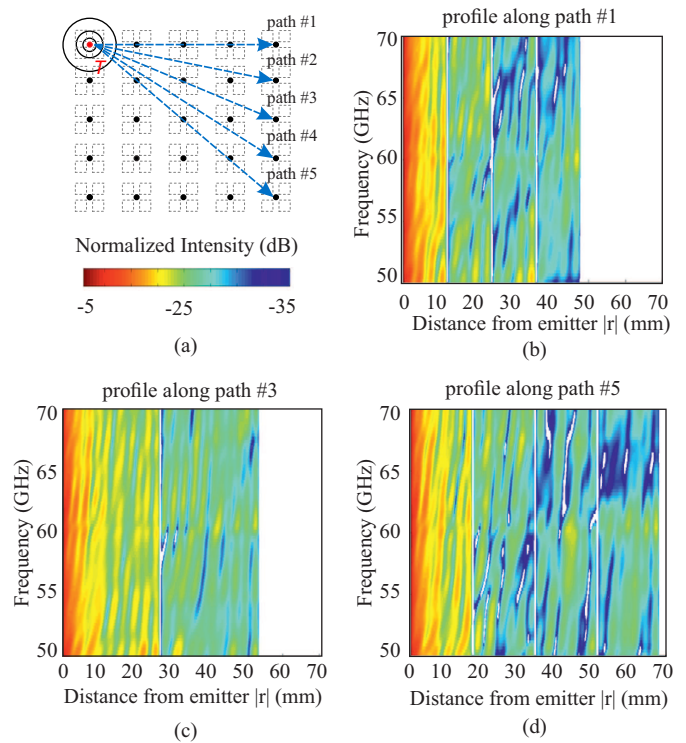}
\vspace{-0.2cm}
\caption{\label{fig:fig_iso_path} (a) A 5x5 grid of antennas where the emitter antenna is no.~1, in the corner. (b,c,d) Power transmission profiles, for frequency $f = $~50 to 70~GHz versus distance from the emitter; the colorbar refers to the power of the total field, emitted plus scattered, calculated along paths (b) \#1, (c) \#3 and (d) \#5, as numbered in panel (a). These paths are selected so that all possible direction are included.}
\vspace{-0.2cm}
\end{figure}

Figure \ref{fig:fig_iso_path} presents the power transmission coefficient (equivalent to path loss, in this 2D approximation) with respect to the frequency and the distance of the emitter in the  frequency range  $f=$~50 to 70~GHz. In particular, in panels (b), (c) and (d) we assume that the element no.~1 radiates and then we calculate the transmission along paths \#1, \#3 and \#5 that connect the position of element no.~1 with elements no.~5, no.~15 and no.~25, respectively, as shown in Fig.~\ref{fig:fig_iso_path}(a). For this calculation we assume that all the elements scatter the incoming field and we do not define a specific receiver. The general trend is that as the distance from the emitter increases the transmission drops. This stands in accordance with the 2$\pi$ angular spread of the emitted energy and the power decay law, i.e. $1/r$. Additionally, we observe that in some positions the transmission drops below $-35$~dB which means that this spots are unreachable for the transmitter. In the power profiles this is indicated by the blank (white) areas. Let us take for example the distribution in path \#1 of Fig. \ref{fig:fig_iso_path}(b); at the positions of the scatterers, $|r| =$ 12, 24 and 36~mm, the field is zero given the presence of the perfect conductors. At frequency $f = 65$~GHz there is a zero-field area between element no.~3 and no.~4, around $|r| = 27$~mm. This kind of intensity mapping can be calculated for every emitter, path and frequency.

\begin{figure*}[ht]\centering
\includegraphics[width=181mm]{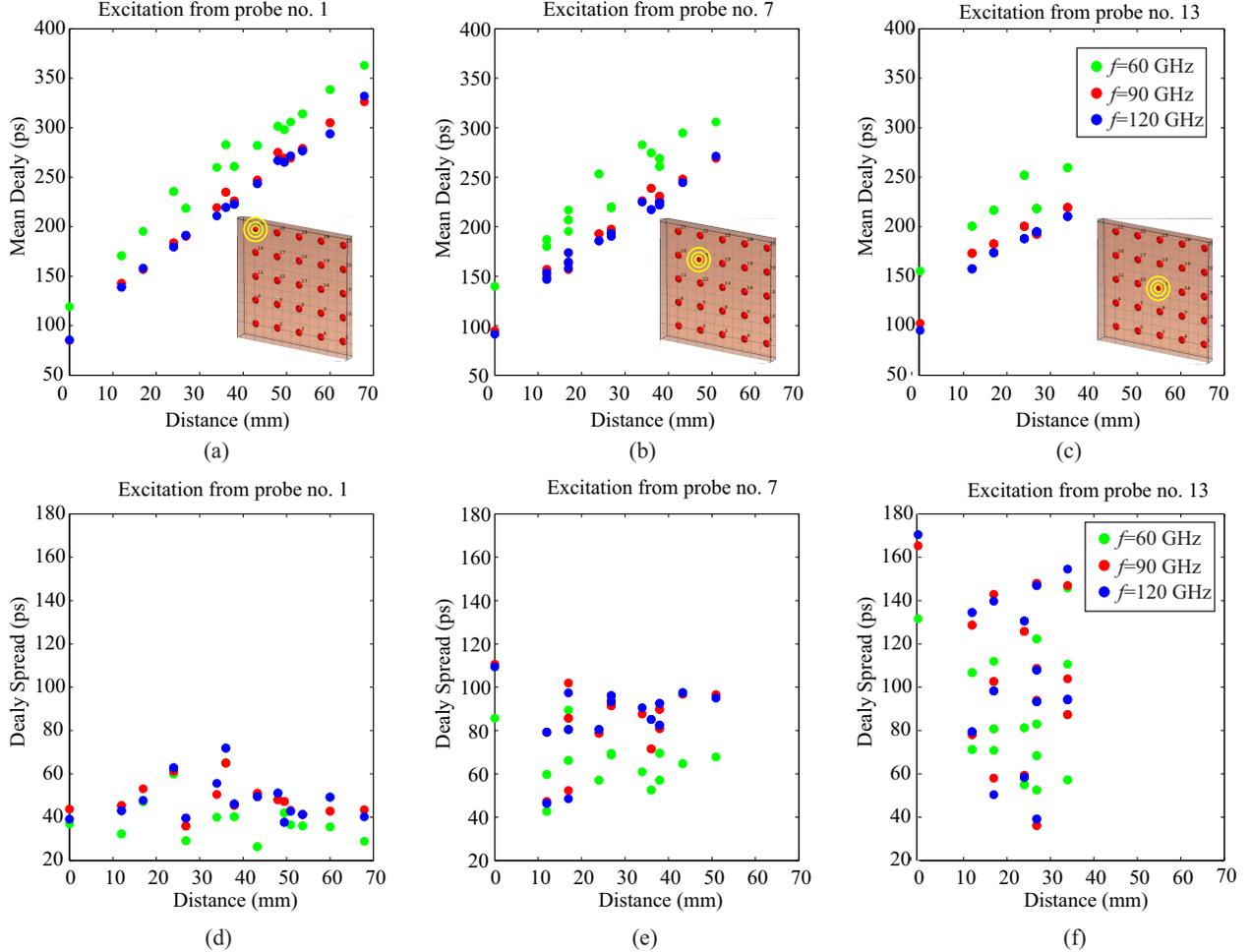}
\vspace{-0.2cm}
\caption{\label{fig:timeC}
Time domain analysis for the dedicated channel for three different frequencies, $f=$~60~GHz (green), 90~GHz (red) and 120~GHz (blue). Top panels present the mean delay assuming that excitation comes from (a) probe no.~1, (b) probe no.~7 and (c) probe no.~13. Insets show the actual 3D simulated structure, the enumeration of the probes in the 5$\times$5 arrangement and the selected radiating probe. Bottom panels show the calculated   delay spread for excitation coming from (d) probe no.~1, (e) probe no.~7 and (f) probe no.~13.
}
\vspace{-0.2cm}
\end{figure*}

Using this 2D qualitative analysis as a guideline, we can select the operation frequency, optimum paths, probe positions, etc., for the actual 3D implementation of the wireless communication channel in the software-defined HSF. The 3D implementation provides the quantitative evaluation of the communication through the time domain analysis and the assessment  of the mean delay and the delay spread.  The analysis is acquired from CST time-domain simulations. We choose to simulate the case of 5$\times$5 probe antennas placed in a rectangular grid of  pitch $D=12$~mm. As mentioned, the grid size is selected with respect to the metasurface operating at 5~GHz. Each 3D antenna  antenna probe is  a single-wire antenna (monopole) placed vertically between the two copper plates. It is implemented by a cylindrical copper rod whose length  $L$ varies according to desired frequency of wireless communication operation. The length of the antenna is  directly connected to the distance of the plates, $d$, in particular $L = d-2 d_{\text{gap}}$, where $d_{\text{gap}}$ is the gap between the antenna and the upper and lower plate and it is assumed much smaller than $L$. The monopole is  excited by  a discrete wire port of impedance equal to $Z_0$~=~50~$\Omega$  that feeds current to the rod and stimulates the radiation or receives radiation from the environment. The port is positioned between the ground (bottom plate) and the  cylindrical rod. The length of the antenna defines the frequency of operation, which is verified in the 3D simulation by the fact that at the resonance frequency where the antenna is matched,  little or no radiation returns to the feeding port ($S_{11}<-20$ dB). The nominal resonant frequency of the antenna is that of the quarter-wave monopole antennas $f=c_0/(4L\sqrt{\varepsilon_\mathrm{r}})$ and for relative small $d_{\text{gap}}$ and $L \approx d$ it is by definition smaller that the cut off frequency of the waveguide; single mode operation is ensured. In what follows we examine three frequency regimes,  $f=$~60, 90 and 120 GHz; the  monopole lengths  where the probe is matched in our 3D implementation are $L=$~0.9, 0.6 and 0.45~mm, receptively. The calculated antenna gain in the azimuth plane is almost unitary (0~dB) and isotropic.

The evaluation of the isolated communication channel in  the time domain  is presented in Fig.~\ref{fig:timeC}. In particular we calculate the  mean delay (top panels of Fig.~\ref{fig:timeC}) and the delay spread (bottom panels of Fig.~\ref{fig:timeC}) for three frequencies, $f=$~60, 90 and 120~GHz and assuming three different non symmetric alternatives  as the radiating probe. In particular we examine the case when probe no.~1 [Fig.~\ref{fig:timeC}(a) and (d)], no.~7 [Fig.~\ref{fig:timeC}(b) and (e)] and no.~13 [Fig.~\ref{fig:timeC}(c) and (f)] feeds the communication. The selection of the three different positions at No.~1, No.~7, and No.~13 in the diagonal of the grid covers a large part of uniquely defined probe and communication paths and pairs in the four fold symmetric 5$\times$5 system.

As a general trend we notice the mean delay  increases linearly with distance which is a feature of the line-of-sight ray transportation which is also observed in the chip communication channel in  Fig. \ref{fig1}(d). This is clearly observed assuming excitation from probe no. 1, shown in Fig.~\ref{fig:timeC}(a), which is also the case that covers the maximum probe-to-probe distance (communication between probes no.~1 and no.~25). The linear trend of the mean delay is less obvious in  Fig.~\ref{fig:timeC}(b) and the case of radiating probe no.~7. In this case we also observe a smaller maximum probe-to-probe distance (communication between probes no.~7 and no.~25) and fewer points on the diagram as a  result of multiple symmetric communication pairs (for example no.7-to-no.2 and no.7-to-no.6). The trend is even less clear in Fig.~\ref{fig:timeC}(c) which is also the point of the highest symmetry in the system. There, the maximum distance becomes very small, actually minimum, (no.~13-no.~25) and we can only observe a general increase of the mean delay with distance. Moreover, in this case the symmetric communication pairs are the most numerous and, therefore, the calculated unique mean delay points are very few. Finally, we observe that in all cases the level of the mean delay is similar for the same distance values. 

The delay spread, on the other hand, exhibits a distinct dependence on the excitation probe. As evident from Fig.~\ref{fig:timeC}(d), (e) and (f) the delay spread is significantly enhanced when the probe is away form the open termination of the structure.  In fact, it is maximized in the case of no.~13 radiating probe which is located at the high symmetry point of the grid, surrounded by multiple densely-packed electromagnetic obstacles which lead to a high number of reflections and increased multipath scattering; this translates into the increase of the delay spread.

\subsection{Discussion}

Communication in the dedicated parallel plate is the channel closest to a wired transmission. The features that attribute the wireless character is the obstacles and some multipath scattering. This channel is not the most efficient one in terms of volume and resources economy since it requires the addition of an extra metallic plate. However, it is a stable solution that allows  the communication between distant nodes in the metasurface tile and can be readily designed for any selected frequency regime. Additionally, the communication channel is practically electromagnetically isolated  from the outer world which means that it is the best solution in terms of security. In fact, a signal coming from the external environment could only couple to the layer through the probe vias which are extremely small thus practically eliminating coupling.

%% file: 5-metasurface.tex
The last investigated option for the intercell communication channel is inside the dielectric layer of the metasurface, between the metallic back plane and the copper patches, as depicted in Fig.~\ref{fig1}(f). The obvious advantage in this approach is the utilization of the metasurface landscape which naturally forms a waveguide channel for communication so that no additional metal plane is required, nor any considerable modification of the HSF architecture or performance. However, this environment hosts a number of characteristics which degrade the channel performance: Firstly, the dielectric medium itself is lossy; secondly, there are gaps between the patches therefore leakage to open space above the metasurface is allowed when the gaps are electrically large; finally, the presence of multiple through-vias (four in each unit cell, connecting the chip with the four patches, see Fig.~\ref{fig1}) imposes unavoidable obstruction and scattering. Moreover, it is noted that this layer has been specifically designed for the operation of the metasurface itself, for example, implementing a tunable absorber for 5~GHz impinging radiation. Consequently, all its geometric and EM parameters, e.g., the dielectric permittivity and thickness, the unit cell and patch sizes etc., have been accurately selected and cannot be modified for the intercell communication. 

For all the reasons outlined, the main parameter available for optimizing the performance in the metasurface channel is the length of the antenna, assumed, in its simplest form, as a monopole connected through the metallic ground plane to the controlling chip. In order to simplify the design and minimize crosstalk and leakage above the copper patches, one monopole antenna is placed in each unit cell, perpendicular to the ground plane and aligned below the center of one of the four patches. This monopole antenna is a copper cylinder of 0.12~mm radius, same as the previous cases, which could be fabricated with a blind via through the ground plane and electroplated. Assuming a small gap of 0.1~mm between the ground plane and the feeding point of the monopole, its length cannot surpass 1.4~mm (when the blind via is drilled), as the thickness of the substrate dielectric is $h=1.575$~mm. Moreover, it must be noted that the thickness of the monopole is expected to affect the performance of the antenna, and it cannot be arbitrarily thin due to fabrication limitations.

Another parameter that could be, \emph{a priori}, freely selected for optimized performance is the intercell communication frequency. In this work, we target 60~GHz, a band of increasing interest for mm-wave communications. However, we note that the natural resonance of the monopole antenna will be closer to 30~GHz; this is due to the presence of the ground plane and the patches plane, which form a quasi quarter-wavelength antenna environment so that $f_\mathrm{res}=c_0/(4h\sqrt{\varepsilon_\mathrm{r}})$, where $h\approx1.5$~mm is the thickness of the dielectric medium (or the maximum length of the antenna) and $\varepsilon_\mathrm{r}=2.2$ its permittivity. Note that increasing the intercell communication frequency above 60~GHz ($\lambda_0=5$~mm) is not a viable option in this metasurface design for two reasons: firstly, because the gap between patches, set at 1.8~mm (approximately one fourth the 60~GHz free-space wavelength), will become electrically large, thus increasing the leakage losses; secondly, because the parallel-plate waveguide formed between the ground and the patches plane will become multimode, as detailed in Section~\ref{sec:dedicated}, which constitutes a suboptimal propagation regime.

Finally, it is worth noting that the guided modes propagating between two penetrable metasurfaces can been formally studied as in Ref.~\cite{Ma2019}. That formulation can be used in the present context, by assuming that the surface impedance of one of the two metasurfaces is zero, thus modeling the uniform ground plane, while the other metasurface is actually the patches plane of the HSF, penetrable by mm-wave radiation.

\subsection{Simulation results}
\begin{figure}[ht]\centering
	\includegraphics[width=85mm]{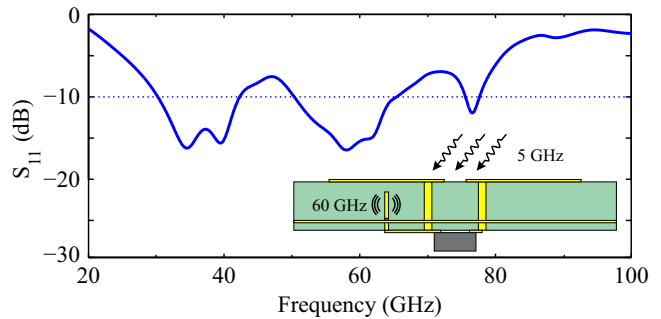}
	\caption{Amplitude of scattering parameter $S_{11}$ for a monopole antenna of $L_\mathrm{ant}=1.2$~mm, at its 50~$\Omega$ feed port marked with red color in the inset, which depicts the cross-sectional side-view of a unit cell for the metasurface-layer intercell communication channel modeling.}
	\label{fig:S11_MSlayerMonopole}
\end{figure}

With all these technical aspects and design choices in mind, and without directly relying on mm-wave matching circuits for the antenna, we assume an S-parameter 50~$\Omega$-reference port between the ground plane and the monopole feed, shown with red in the base of the antenna in the inset of Fig.~\ref{fig:S11_MSlayerMonopole}. Full-wave numerical simulations are conducted using CST Microwave Studio throughout this Section. We numerically calculate the monopole length where the antenna resonates or, similarly, where the amplitude $|S_{11}|$ parameter is minimized. A local optimal value for the monopole length was found near $L_\mathrm{ant}=1.2$~mm, which leads to the $|S_{11}|$ spectrum depicted in Fig.~\ref{fig:S11_MSlayerMonopole}, opening two well matched bands near 35 and 60~GHz. It must be noted that a proper resonance of the monopole, where $\mathrm{Im}\{Z_\mathrm{ant}\}=0$ (zero reactance in the antenna input impedance), cannot be attained at 60~GHz for the given metasurface environment, so the $|S_{11}|\approx-15$~dB value is deemed sufficient. The calculated antenna gain in the azimuth plane is approximately unitary (0~dB); a few narrow-beamwidth cancelling directions can be found towards the center of the unit cell where the four through vias act as reflectors, with extra gain in the opposite direction. The antenna is polarized primarily perpendicular to the ground plane. Concluding the design of the antenna, we verified that the coupling between the antenna feed port and both the chip ports and the Floquet port at the receiving side of the metasurface is negligible at its operation frequency (5~GHz); this eliminates cross-talk between the intercell communication channel and the main metasurface functionality.

\begin{figure}[ht]\centering
	\includegraphics[width=85mm]{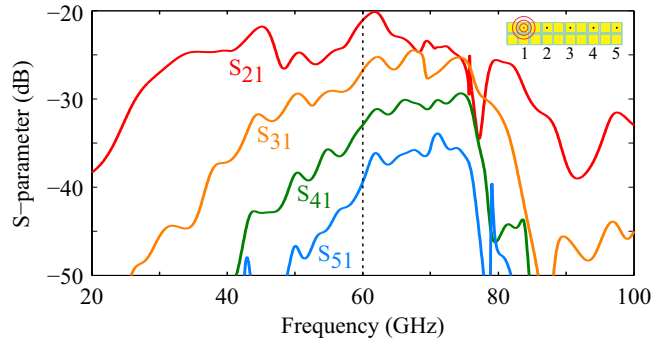}
	\caption{Transmission spectra in terms of S-parameter amplitude (in dB) for a $5\times1$ arrangement of unit cells where the transmitting antenna is in the first cell, as shown in the inset.}
	\label{fig:Sm1_5x1_grid}
\end{figure}

Moving on to the frequency domain characterization of the intercell communication channel, we consider five unit cells in a row with a transmitting antenna placed in the first cell, and numerically calculate the transmission to the other cells, in terms of the scattering parameter amplitude $|S_{i1}|$, where $i=2,3,4,5$. The resulting spectra are depicted in Fig.~\ref{fig:Sm1_5x1_grid}. The transmission to the adjacent neighbouring cell in 60~GHz is approximately $|S_{21}|\approx-20$~dB and another 6~dB of path loss are accumulated with each added cell of lateral width (pitch) equal to 12~mm. The highest transmission band is in all cases near 60~GHz, well suited with the band where the antenna is well matched (exhibits low values of $|S_{11}|$), and with a bandwidth decreasing as the unit-cell distance increases. A sharp drop in transmission takes place close to 80~GHz which is attributed to leakage between the copper patches, outside the metasurface dielectric.

\begin{figure}[ht]\centering
	\includegraphics[width=85mm]{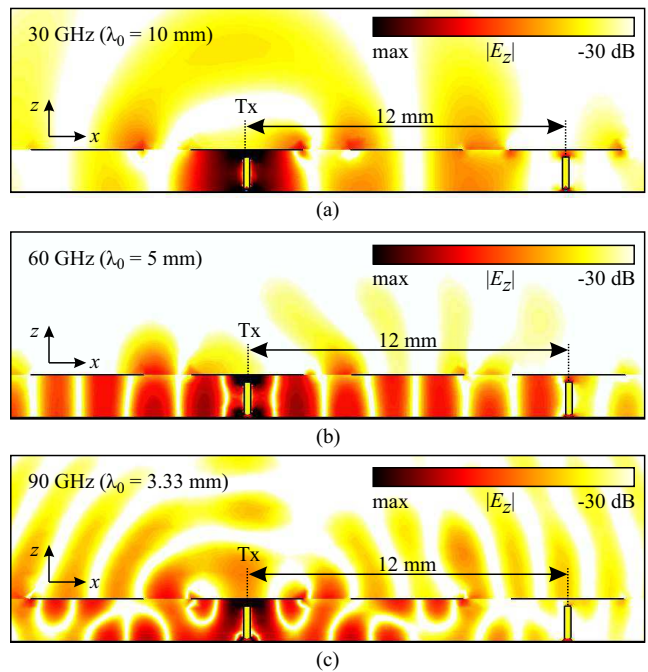}
	\caption{Field plots of the wave radiated from the 1.2~mm tall monopole antenna inside the metasurface layer at (a) 30, (b) 60 and (c) 90 GHz. The plots correspond to the amplitude of the vertically polarized E-field component, in the vertical cross-section plane passing through the axes of two neighbouring monopole antennas.}
	\label{fig:FieldPlots_MSLayer}
\end{figure}

In order to visualize the performance of the monopole antenna, we depict the radiated wave from one unit cell to its neighbouring one, at three different frequencies: 30, 60 and 90 GHz, in Fig.~\ref{fig:FieldPlots_MSLayer}. The plots correspond to the amplitude of the E-field component polarized perpendicularly to the ground plane and the patches, in a vertical cross-section including two neighbouring unit cell antennas. Evidently, the optimal performance is at the design frequency of 60~GHz, where the field amplitude reaching the neighbouring cell is maximized. In both 30 and 90~GHz, the antenna radiates strongly outside the metasurface layer, which leads to the lower transmission values ($|S_{21}|$) observed in Fig.~\ref{fig:Sm1_5x1_grid}. This behaviour can be qualitatively explained by considering the waveguide formed between the ground plane and the patches plane \cite{Ma2019}, whose cut-off frequency is approximately 67~GHz as described in Section \ref{sec:dedicated}; optimal confinement in the metasurface layer, and thus highest transmission, is attained for the frequency right below the cut-off, whereas confinement is reduced far below the cut-off (noting, also, that the patches plane is penetrable) and multimode regime is entered above the cut-off.

\begin{figure}[ht]\centering
	\includegraphics[width=85mm]{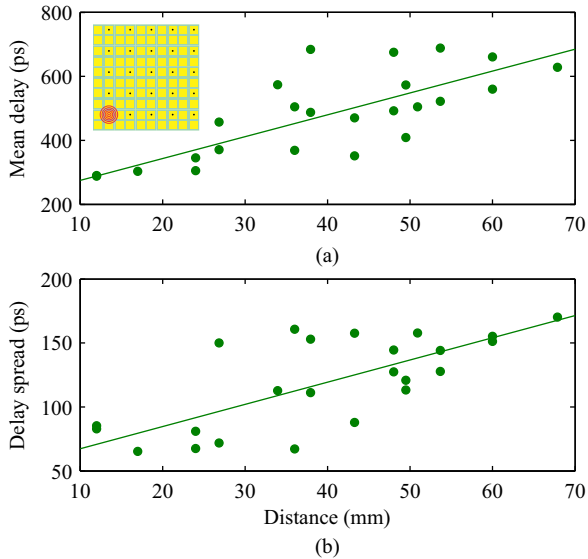}
	\caption{(a) Mean delay and (b) RMS delay spread extracted with time-domain simulation of a $5\times5$ arrangement of unit cells where the 60~GHz transmitting antenna is in one of the corner cells.}
	\label{fig:TD_5x5_grid}
\end{figure}

Having quantified the path loss in the metasurface intercell communication channel, we conclude our analysis with the time domain metrics that offer an estimate of the coherence bandwidth. We assume a $5\times5$ arrangement of unit cells where the transmitting antenna is in one of the corner cells, similar to Fig.~\ref{fig:fig_iso_pairs}(a). The mean delay, $\overline{\tau_{ij}}$ and RMS delay spread, $\tau_{\text{rms}}^{(i,j)}$, are acquired from CST time-domain simulations post-processed with Eqs.~(\ref{eq:meanDelay}) and (\ref{eq:DelaySpread}). The transmitted pulse bandwidth used for the calculations occupies a spectrum of $\pm20\%$ around the central frequency, 60~GHz. The results are depicted in Fig.~\ref{fig:TD_5x5_grid}, where a 6~ps/mm and a 1.5~ps/mm trendline emerges for the mean delay and delay spread, respectively. The large deviation in the values of mean delay is in contrast with what was calculated in the previous communication channels, which is attributed to the structured geometry of the metasurface and multi-scatterer environment (four through-vias per unit cell). For the same reason, the maximum delay spread is rather large, exceeding 150~ps, which leads to a coherence bandwidth of only 7~GHz ($10\%$ of the operating frequency). Note that we limited the time-domain study to the 60~GHz band, as higher frequencies suffer from poor antenna matching and high path loss, as we evidenced in the frequency domain analysis.

\subsection{Discussion}

Summarizing the analysis and design of the intercell communication in the metasurface layer, we have chosen to work with prescribed geometric and EM properties for the environment and design a simple low-cost monopole 60~GHz antenna in a perturbation approach, i.e., with the aim of minimizing the effect on the main metasurface operation, which is at 5~GHz. We then proceeded to evaluate the metrics of the channel model, namely the path loss and delay spread. Allowing a small amount of perturbation to the metasurface parameters, e.g. the dielectric thickness or the patch width, and/or freely choosing the intercell communication frequency band as fits best, and/or invoking microwave circuits to match $|S_{11}|$ at will, can lead to improved designs \cite{Tasolamprou2018}. The main drawback in these approaches is the custom/non-standard ASICs (chips) required to compensate and even-out performance perturbations. Finally, arguably the optimal approach would be to `co-design' the metasurface layer with both the main functionality and the intercell communication aspects factored in, right from the start; this could potentially lead to overall optimized performance at the cost of more resources and design iterations.